\documentclass[
aps,
nofootinbib,
prd,
superscriptaddress,
tightenlines,
notitlepage,
twocolumn,
showpacs,
floatfix
]{revtex4-2}

\usepackage{amsmath}
\usepackage{latexsym}
\usepackage{amsfonts}
\usepackage{graphicx}
\usepackage{mathrsfs}
\usepackage{epstopdf}
\usepackage{subfigure}
\usepackage[utf8]{inputenc}
\usepackage{CJK}
\usepackage{float}
\usepackage{color}
\usepackage{hyperref}
\usepackage{diagbox}
\usepackage{xcolor}
\usepackage{bm}
\usepackage{lipsum}
\hypersetup{
    colorlinks=true,
    linkcolor=red,
    citecolor=blue,
}
\usepackage{ulem}

\def\al{\alpha}

\def\de{\delta}
\def\ep{\epsilon}

\def\et{\eta}
\def\th{\theta}

\def\ka{\kappa}

\def\ta{\tau}

\def\ph{\phi}

\def\om{\omega}

\def\frac#1#2{{\textstyle{{#1}\over {#2}}}}

\def\lsim{\mathrel{\rlap{\lower4pt\hbox{\hskip1pt$\sim$}}
    \raise1pt\hbox{$<$}}}
\def\gsim{\mathrel{\rlap{\lower4pt\hbox{\hskip1pt$\sim$}}
    \raise1pt\hbox{$>$}}}
\def\sqr#1#2{{\vcenter{\vbox{\hrule height.#2pt
         \hbox{\vrule width.#2pt height#1pt \kern#1pt
         \vrule width.#2pt}
         \hrule height.#2pt}}}}

\def\lrpartial{\raise 1pt\hbox{$\stackrel\leftrightarrow\partial$}}

\def\part2{\partial_\alpha \partial^\alpha}

\def\xx'{|\vec x -\vec x'|}

\def\b2{b^\al b_\al}

\newcommand{\bw}{\begin{widetext}}
\newcommand{\ew}{\end{widetext}}
\newcommand{\beq}{\begin{equation}}
\newcommand{\eeq}{\end{equation}}
\newcommand{\bea}{\begin{eqnarray}}
\newcommand{\eea}{\end{eqnarray}}
\newcommand{\bit}{\begin{itemize}}
\newcommand{\eit}{\end{itemize}}
\newcommand{\rf}[1]{(\ref{#1})}

\begin{document}

\title{Probing vector hair of black holes with extreme mass ratio inspirals}

\author{Dicong Liang}
\email{dcliang@pku.edu.cn}
\affiliation{Kavli Institute for Astronomy and Astrophysics, Peking University, Beijing
100871, China}

\author{Rui Xu}
\email[Corresponding author: ]{xuru@pku.edu.cn}
\affiliation{Kavli Institute for Astronomy and Astrophysics, Peking University, Beijing
100871, China}

\author{Zhan-Feng Mai}
\affiliation{Kavli Institute for Astronomy and Astrophysics, Peking University, Beijing
100871, China}

\author{Lijing Shao}
\email[Corresponding author: ]{lshao@pku.edu.cn}
\affiliation{Kavli Institute for Astronomy and Astrophysics, Peking University, Beijing 100871, China}
\affiliation{National Astronomical Observatories, Chinese Academy of Sciences, Beijing 100012, China}

\begin{abstract}
The bumblebee gravity model, with a vector field nonminimally coupled to gravity, is a natural extension of the Einstein-Maxwell theory. In this theory, a black hole can carry a vector hair, making the metric deviate from the Schwarzschild metric.
To investigate the detectability of the vector hair, we consider an Extreme Mass Ratio Inspiral (EMRI) system, where 
a stellar-mass black hole inspiraling into a supermassive black hole.
We find that, with a one-year observation of an EMRI by a space-based gravitational-wave detector, we can probe the vector charge as small as $Q\sim 10^{-3}$ in the bumblebee gravity model, which is about three orders of magnitude tighter comparing to current EHT observations.
\end{abstract}
\maketitle

\allowdisplaybreaks 

\section{Introduction}

It is believed that there is a population of supermassive black holes (SMBHs) with masses in the range $10^5$–$10^9M_\odot$, residing at the center of galaxies \cite{lynden1971quasars,Soltan:1982vf,Kormendy:1995er,Gultekin:2009qn,Genzel:2010zy}. 
Electromagnetic observations with the Event Horizon Telescope (EHT) have provided convincing evidence for the presence of SMBHs in the galaxy M87 and the Milky Way \cite{EventHorizonTelescope:2019dse,EventHorizonTelescope:2022wkp}.
With the EHT images of the black holes (BHs), we can not only know more about the properties of the BHs and their surrounding environment, but also use them to test general relativity (GR)
\cite{EventHorizonTelescope:2019ggy,EventHorizonTelescope:2019pgp,EventHorizonTelescope:2020qrl,EventHorizonTelescope:2021srq,EventHorizonTelescope:2021dqv,EventHorizonTelescope:2022xqj,Broderick:2013rlq,Johnson:2015iwg,Bambi:2019tjh,Davoudiasl:2019nlo}.
In the near future, we can use gravitational waves (GWs), as another window to observe the SMBHs with space-based detectors like the Laser Interferometer Space Antenna (LISA) \cite{Danzmann:1997hm,LISA:2017pwj,Bayle:2022hvs}, Taiji \cite{Hu:2017mde} and TianQin \cite{TianQin:2015yph,Gong:2021gvw}.

An extreme mass ratio inspiral (EMRI) system, comprised of a SMBH and a stellar-mass compact object, provides unprecedented opportunity to probe the nature of the SMBH \cite{Amaro-Seoane:2007osp}. 
Such a system is expected to complete around $10^4$ to $10^5$ cycles in the detector sensitive band, thus allowing exquisite measurements of system parameters \cite{Babak:2017tow}. 
EMRI not only can be used to test fundamental physics \cite{LISA:2022kgy,Barack:2006pq,Barausse:2020rsu,Zi:2021pdp,Rahman:2022fay,Lukes-Gerakopoulos:2021ybx,Destounis:2020kss,Destounis:2021mqv,Destounis:2021rko,Maggio:2021uge,Deich:2022vna}, 
but also has many astronomical and cosmological applications \cite{Berry:2019wgg,Laghi:2021pqk}. The astrophysical environment around the SMBH can be studied with the GWs from the EMRI system \cite{Dai:2021olt,Dai:2023cft,Li:2021pxf,Cardoso:2022whc,Becker:2022wlo,Rahman:2021eay,Barausse:2006vt,Macedo:2013qea,Barausse:2014tra,Cardoso:2019rou,Toubiana:2020drf,Speri:2022upm,Sberna:2022qbn,Polcar:2022bwv,Cardoso:2021wlq}

In a large class of scalar-tensor theories, due to the existence of the scalar field, there is extra dipolar and monopolar radiation for binary systems \cite{Eardley1975,Will:1977wq,Will:1989sk,Damour:1998jk,Barausse:2015wia,Doneva:2017bvd,Xu:2021kfh}. The extra radiation channel leads to dephasing effects in GWs.
Thus, EMRIs can be used to probe scalar fields \cite{Maselli:2020zgv,Maselli:2021men,Jiang:2021htl,Guo:2022euk,Zhang:2022rfr,Barsanti:2022ana,Barsanti:2022vvl}. 
On the other hand, if the smaller compact object carries electric charge, during its inspiral into the SMBH, the system also emits electromagnetic waves. Using the dephasing effects to probe electric charge of the small compact object was studied by \citet{Zhang:2022hbt}, and then it has been extended to vector charge in Ref.~\cite{Zhang:2023vok}.
\citet{Burton:2020wnj} considered an EMRI system with two charged BHs, i.e. the central SMBH is a Reissner-Nordstr\"om BH.
Furthermore, the inspiral of a charged small compact object into a  charged rotating SMBH, i.e.a\ a Kerr-Newman BH, was studied in Ref.~\cite{Zi:2022hcc}.

In this paper, we investigate EMRIs in a vector-tensor theory called the bumblebee gravity model.
The action of the model is given by \cite{Kostelecky:2003fs}
\begin{align}
\label{bumblebeeaction}
S=\int \sqrt{-g} d^4x \bigg[& \frac{1}{2\kappa} (R+\xi B^\mu B^\nu R_{\mu\nu}) -\frac{1}{4} B_{\mu\nu}B^{\mu\nu}
\nonumber \\
&  -V(B^\mu B_\mu \pm b^2) \bigg] +S_m ,
\end{align}
where $B_{\mu\nu}=D_\mu B_\nu -D_\nu B_\mu$,
$\kappa=8\pi G$, $\xi$ is the coupling constant, $V$ is the potential for the vector field and $b^2$ is a real positive constant. 
The bumblebee gravity model has been studied widely in the literature \cite{Bailey:2006fd,Bluhm:2004ep,Bluhm:2007bd,Bertolami:2005bh,Casana:2017jkc,Xu:2022frb,Ovgun:2018xys,Bluhm:2008yt,Liang:2022hxd,Liu:2022dcn,Maluf:2021lwh,Khodadi:2022mzt}.
It can be considered as a natural extension of the Einstein-Maxwell theory, with the vector field nonminimally coupled to gravity.
In our previous study \cite{Xu:2022frb}, we found a novel vacuum solution for the bumblebee gravity model, where the BH can carry a vector hair.
However, the parameter space remained largely unexcluded with current EHT observations of the shadows of SMBHs, i.e.\ Sgr A* and M87 \cite{Xu:2022frb}. As we find in this paper, with the observation of EMRIs, we can probe the vector charge of the SMBH to a level as small as $Q\sim \mathcal{O}(10^{-3})$, which is about three orders of magnitude smaller than that with the EHT.

The paper is organized as follows. First, we give a brief overview of the hairy BH solution in the bumblebee model in Sec.~\ref{BHsolution}.
Next, we compute the circular orbits of EMRIs in both GR and bumblebee gravity model in Sec.~\ref{orbit}.
Then, we compare the waveforms in the two gravity theories and calculate the faithfulness to show the detectability of the vector hair in Sec.~\ref{waveformcom}.
Last, we have a final discussion in Sec.~\ref{discussion}.

We use the geometrized unit system where $G = c = 1$ when deriving the BH solution. The sign convention of the metric is $(-, +, +, +)$. When comparing with the Reissner-Nordstr\"om solution, we also set the free-space permittivity $\epsilon_0$ to be unity.

\section{Bumblebee black hole}
\label{BHsolution}

As we discussed,  the bumblebee gravity model can be regarded as a generalization of the Einstein-Maxwell theory. We therefore expect that spherical black hole solutions with vector hair in the bumblebee model extend the Reissner-Nordstr\"om solution. The details of construction of the general static BH solutions in the bumblebee model can be found in Ref.~\cite{Xu:2022frb}, and we give a brief overview here.
To construct the BH solutions, we first use the metric ansatz.
\bea
ds^2 =  -e^{2\nu} dt^2 + e^{2\mu} dr^2 + r^2 \left( d\th^2 + \sin^2\th d\ph^2 \right),
\label{ssmetric}
\eea  
and assume the bumblebee field to be $b_\mu = \left( b_t, \, 0, \, 0, \, 0\right)$, where $\nu, \, \mu$, and $b_t$ are functions of $r$ to be solved from the field equations. 
A more general assumption for the bumblebee field with spherical symmetry should include a radial component $b_r$. 
Such solutions are indeed found in Ref.~\cite{Xu:2022frb}, but they cannot recover the Reissner-Nordstr\"om solution when $\xi$ is zero. 
In fact, they only exist for $\xi \ne 0$. So they are extraneous solutions indubitably caused by the nonminimal coupling. 
Though beyond the scope of the present work, investigating the ability to test these solutions with the space-based detectors would be worth considering in a future study. Besides that, we do not consider nonspherical solutions in this work. 
For a stable vacuum, we further require that the potential $V$ and its derivative with respect to the bumblebee field vanish. 
It turns out that the field equations reduce to three ordinary differential equations for $\nu, \, \mu$, and $b_t$, 
\bw
\bea
0 &=& \frac{e^{2 \mu}-2 r \nu'-1}{r^2} - e^{-2 \nu} \left( \frac{\ka}{2} b_t^{\prime\,2} - \xi b_t b_t^{\prime} \nu' + \xi b_t^2 \nu'^2\right) ,
\nonumber \\
0 &=& r^2 e^{-2 \mu} \left( \mu'\nu'+ \frac{1}{r}\mu'-\frac{1}{r}\nu'- \nu''- \nu'^2\right) + r^2 e^{-2 (\mu+\nu)} \left( \frac{\ka}{2} b_t^{\prime\,2} - \xi b_t b_t^{\prime} \nu' + \xi b_t^2 \nu'^2 \right) ,
\nonumber \\
0 &=& b_t^{\prime\prime}- b_t^{\prime} \left( \mu'+ \nu'-\frac{2}{r}\right)+ \frac{\xi}{\ka} b_t \left( \mu' \nu' - \frac{2}{r}\nu' - \nu'' - \nu'^2 \right) ,
\label{odes} 
\eea 
\ew
where the prime denotes the derivative with respect to $r$.

We cannot find analytical solutions to Eq.~\rf{odes} for a general coupling constant $\xi$, so a numerical approach is taken to obtain the solutions. As we are looking for solutions of spherical BHs, at the radius of the event horizon $r_h$, we have $g_{tt} = -e^{2\nu} \rightarrow 0$ and $g_{rr} = e^{2\mu} \rightarrow \infty$. Therefore, we assume the expansions
\bea
&& g_{tt} = - \left( N_{11} \de + N_{12} \de^2 + N_{13} \de^3 + ... \right) ,
\nonumber \\
&& g_{rr} = \frac{1}{\de} \left( M_{10} + M_{11} \de + M_{12} \de^2 + ... \right) , 
\nonumber \\
&& b_t  = L_{10} + L_{11} \de + L_{12} \de^2 + ... ,
\label{hexp}
\eea 
where $\de \equiv r-r_h$, for the metric functions and the nonzero bumblebee component near the horizon. Note that the notations for the expansion coefficients follow Ref.~\cite{Xu:2022frb}. Substituting the expansions into Eq.~\rf{odes}, we find recurrence relations for the undetermined coefficients. The first few are
\bw
\bea
M_{10} &=& r_h + \left( \frac{\ka}{2} - \frac{\xi}{4} \right) \frac{L_{11}^2 r_h^2}{N_{11}} ,
\nonumber \\
L_{10} &=& 0 ,
\nonumber \\
M_{11} &=& 1 - \left( \frac{\ka^2}{4}-\ka \xi+\frac{13 \xi^2}{16}-\frac{3 \xi^3}{16 \ka} \right) \frac{L_{11}^4 r_h^2}{N_{11}^2}
 + \left( \frac{3 \ka^2 \xi}{8}-\frac{9 \ka \xi^2}{16}+\frac{9 \xi^3}{32}-\frac{3 \xi^4}{64 \ka} \right) \frac{L_{11}^6 r_h^3}{N_{11}^3},
\nonumber \\
M_{12} &=& \left( \frac{\ka^2}{4}-\ka \xi+\frac{13 \xi^2}{16}-\frac{3 \xi^3}{16 \ka}\right) \frac{L_{11}^4 r_h}{N_{11}^2}
 + \left( \frac{\ka^3}{8}-\frac{19 \ka^2 \xi}{12}+\frac{105 \ka \xi^2}{32}-\frac{173 \xi^3}{64}+\frac{191 \xi^4}{192 \ka}-\frac{35 \xi^5}{256 \ka^2} \right) \frac{ L_{11}^6 r_h^2}{ N_{11}^3}
\nonumber \\
&& - \left( \frac{49 \ka^3 \xi}{96}-\frac{433 \ka^2 \xi^2 }{192 }+\frac{207 \ka \xi^3 }{64 }-\frac{809 \xi^4 }{384 }+\frac{997 \xi^5 }{1536 \ka }-\frac{79 \xi^6 }{1024 \ka^2 }\right) \frac{L_{11}^8 r_h^3}{N_{11}^4}
\nonumber \\
&& + \left( \frac{11 \ka^3 \xi^2 }{32 }-\frac{55 \ka^2 \xi^3 }{64 }+\frac{55 \ka \xi^4 }{64 }-\frac{55 \xi^5 }{128}+\frac{55 \xi^6}{512 \ka }-\frac{11 \xi^7 }{1024 \ka^2 } \right) \frac{L_{11}^{10} r_h^4}{N_{11}^5} ,
\nonumber \\
N_{12} &=& -\frac{N_{11}}{r_h} + \left( \frac{ \ka}{2}-\frac{ \xi}{4}\right) L_{11}^2 + \left( \frac{ \ka \xi}{4 }-\frac{ \xi^2}{4} + \frac{ \xi^3}{16 \ka}\right) \frac{L_{11}^4 r_h}{N_{11}} ,
\nonumber \\ 
L_{12} &=& -\frac{L_{11}}{r_h} + \left( \frac{ \xi}{4 } -\frac{ \xi^2}{8 \ka} \right) \frac{L_{11}^3}{N_{11}} + \left( \frac{ \ka\xi}{4 }-\frac{ \xi^2}{4 } + \frac{ \xi^3}{16 \ka} \right) \frac{L_{11}^5 r_h}{N_{11}^2}   .
\eea
\ew
With the expansions in Eq.~\rf{hexp}, and after given $\xi,\, r_h,\, N_{11},$ and $L_{11}$, numerical integrations for Eq.~\rf{odes} can be initialized at a radius slightly larger than $r_h$ and outward. Integrating to a radius numerically large enough, the mass $m$ and the bumblebee charge $Q$ can be extracted from the solutions of $\mu$ and $b_t$ using
\bea
m = -\lim\limits_{r\rightarrow \infty} r^2 \mu' ,\quad Q = -\sqrt{ \frac{\ka}{2} } \lim\limits_{r\rightarrow \infty} r^2 b_t' .
\eea 
Note that the constant $\sqrt{\ka/2}$ is used for the solutions to recover the Reissner-Nordstr\"om solution with
\bea
&& \nu = -\mu = \frac{1}{2} \ln{\left(1-\frac{2 m}{r} + \frac{Q^2}{r^2} \right)}, 
\nonumber \\
&& b_t =  \frac{1}{ \sqrt{4\pi} } \frac{Q}{ r} ,
\eea
when $\xi = 0$. 
The charge can be positive or negative, which depends on the sign of $b_t$ we choose. Since the modifications on the Schwarzschild metric are the same for $+Q$ and $-Q$, we only consider $Q\geq0$ in this paper. 

This BH solution has a novel feature, that is, when $\xi=2\kappa$, it reduces to a stealth Schwarzschild solution, i.e.\ a Schwarzschild metric, accompanied with a nontrivial and regular vector field.
Stealth Schwarzschild solutions can also be found in other vector-tensor theories and scalar-tensor theories \cite{Cisterna:2016nwq,Heisenberg:2017hwb,Heisenberg:2017xda,Babichev:2013cya,Babichev:2017guv,Charmousis:2019vnf}.
The coupling constant $\xi$ represents for the coupling strength between the vector field and the curvature, we expect it to be very small, since no significant GR-violating evidence has been found so far. We want to investigate if we can probe the small modification on the Schwarzschild spacetime with EMRIs, thus we consider the value of $\xi$ to be around $2\kappa$ in this paper.  

\begin{figure}
 \includegraphics[width=\linewidth]{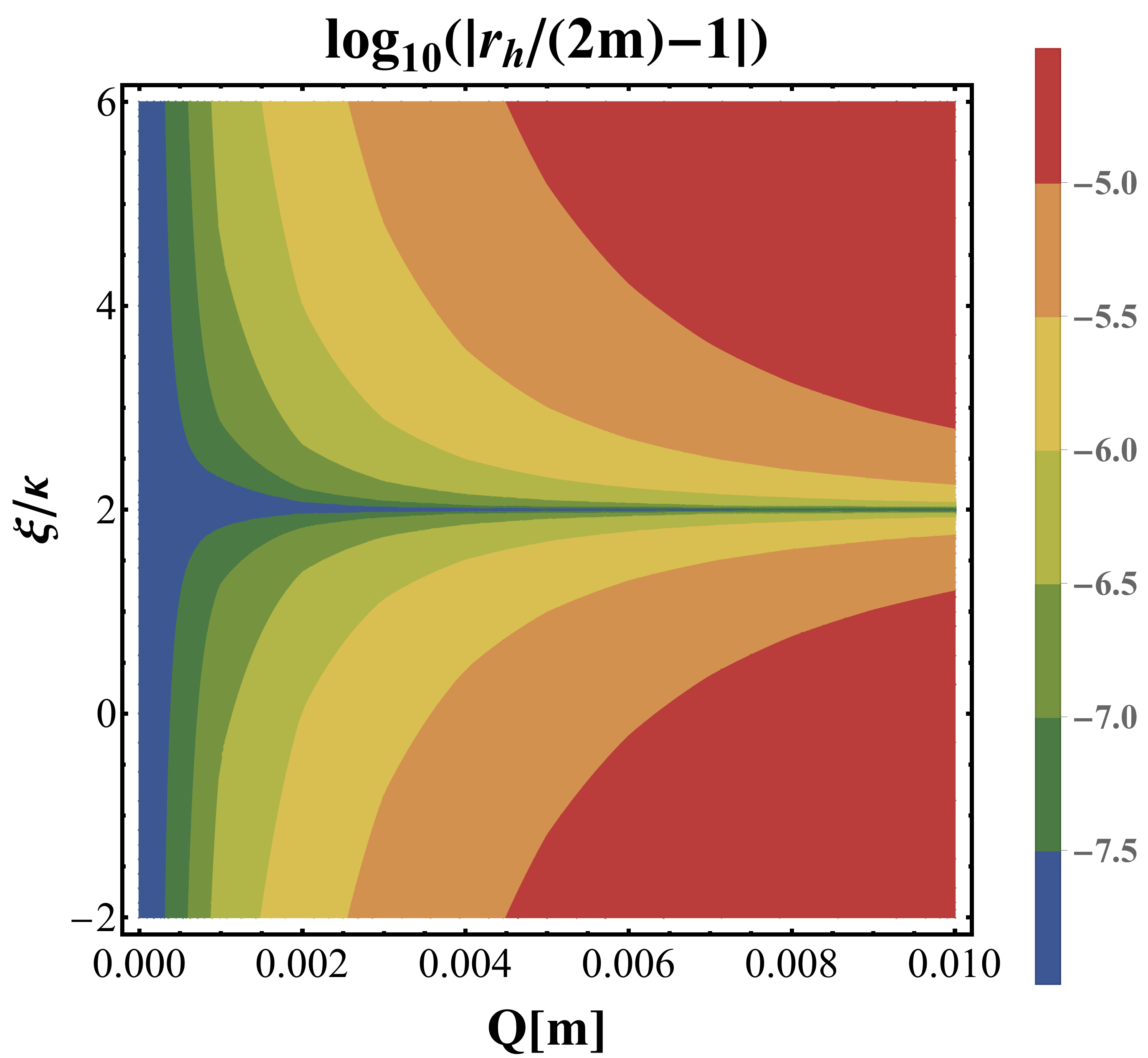}
 \caption{Relative deviation of the radius of horizon $r_h$ from $2m$ in the bumblebee model as a function of  vector charge $Q$ and coupling constant $\xi$. We calculate the absolute value of the relative deviation in the logarithm scale.}
\label{rh}
\end{figure}

When $\xi\neq 2\kappa$, the deviation from the Schwarzschild metric in a hairy bumblebee BH grows with the vector charge. The deviation is also reflected in the radius of the horizon \cite{Xu:2022frb}. 
It is widely known that the radius of the horizon is $r_h=2 m$ for a Schwarzschild BH in the geometric unit. 
When the BH carries a small vector charge, the radius $r_h$ will deviate from $2m$ slightly.
Specifically, we have $r_h>2m$ when $\xi>2\kappa$ and $r_h<2m$ when $\xi<2\kappa$.
Figure~\ref{rh} shows the horizon radius $r_h$ with respect to $Q$ in unit of the BH mass $m$ for different choices of the coupling constant $\xi$. 
Similar to Einstein-Maxwell theory, the vector charge cannot be arbitrarily large. There is a maximum charge that a BH can carry for different $\xi$ (cf.\ Fig.\ 5 in Ref.~\cite{Xu:2022frb}).
The larger the vector charge is, the larger deviation from the Schwarzschild spacetime. 
Since we want to explore how small vector charge we can detect with EMRIs, we focus on a small relevant region in Fig.~\ref{rh}.  
When there is no vector charge or $\xi=2\kappa$, the metric reduces to the Schwarzschild metric, namely $r_h=2 m$. 
In our numerical solution, we cannot achieve this exactly due to numerical errors. We find a deviation of $r_h/(2m)-1\sim \mathcal{O}(10^{-9})$ when setting $Q$ to $0$ or $\xi$ to $2\ka$ numerically. 
That indicates the magnitude of the error in the numerical solutions. 

\section{Adiabatic inspiral}
\label{orbit}

In an EMRI system, the mass of the smaller compact object is much smaller than that of the SMBH, thus it can be treated as perturbation in the background spacetime produced by the SMBH. 
Besides, the radiation reaction takes place over a time scale much longer than the orbital period, i.e.\ the adiabatic approximation is valid \cite{Cutler:1994pb}.
Thus, the motion of the small compact object is approximated to be geodesic over several orbital periods. Then we can calculate the time-averaged loss of the energy and the changing rates of  orbital parameters. Finally we can obtain the slow, secular evolution of the orbit \cite{Cutler:1994pb,Babak:2006uv,Chua:2017ujo}.

Differently from the electromagnetism, we do not consider the coupling between the vector field and the matter field, which means that we have the variation $\delta S_m/\delta B_\mu=0$. 
Then, the perturbation of the vector field $\tilde{B}^\mu$ satisfies the following equation of motion
\beq
D^\mu (D_\mu \tilde{B}_\nu -D_\nu \tilde{B}_\mu) +\frac{\xi}{\kappa} \tilde{B}^\mu \bar{R}_{\mu\nu}  = -\frac{\xi}{\kappa} b^\mu \tilde{R}_{\mu\nu}.
\eeq
Here, $\bar{R}_{\mu\nu}$ is the background curvature produced by the hairy SMBH and the source $\tilde{R}_{\mu\nu}$ is produced by the GWs.
Thus, we do not have the electriclike current and there is no dipole radiation for the bumblebee field.
When we consider a small vector field $|b^\mu|\ll 1$, the source of the perturbation $\tilde{B}^\mu$ is suppressed, compared to the tensor modes of GWs.
As indicated in Ref.~\cite{Liang:2022hxd}, the extra modes of GWs couple to the perturbations of the bumblebee field, and is suppressed by $\xi$. 
Thus in this paper, we will drop all their contribution to the energy loss and only consider the tensor modes. 
In other words, we consider that the orbit of an EMRI in the bumblebee model differs from that in GR mainly due to the difference of the background metric produced by the central SMBH. 
The smaller BH will be used as a probe to measure the deviation of the metric from the Schwarzschild metric.
Taking into consideration of the extra modes in future work would be necessary for more accurately calculating the orbital evolution and building the waveform. We will have more discussion on this issue in Sec.~\ref{discussion}.

We consider adiabatically decaying circular orbits, where the energy carried away by the tensor modes at the leading order is given by \cite{Poisson:2014}
\bea
\frac{dE}{dt}= - \frac{32}{5} \left( \et M \om^3 r^2 \right)^2 .
\label{decay1}
\eea
Here, $M = m_1 + m_2$ is the total mass of the two BHs, $\et = m_1m_2/M^2$ is the symmetric mass ratio, and $\om$ and $r$ are the angular velocity and the radius of the circular orbit. In the spherical spacetime represented by Eq.~\rf{ssmetric}, the timelike geodesic equation has the first integrals
\bea
&& \frac{dt}{d\ta} = \ep e^{-2\nu} , 
\nonumber \\
&& \left( \frac{dr}{d\ta} \right)^2 = e^{-2\mu } \left(  \ep^2 e^{-2\nu} - \frac{l^2}{ r^2 } - 1  \right) ,
\nonumber \\
&& \frac{d\ph}{d\ta} = \frac{l}{r^2} ,
\eea  
where $\ta$ is the proper time along the geodesic, $\ep$ is the specific energy constant, and $l$ is the specific angular momentum constant. We have set the orbit in the $\th = \pi/2$ plane utilizing the spherical symmetry. Together with the equation for the radial acceleration
\bea
\frac{d^2r}{d\ta^2} = e^{-2\mu } \left[ -\ep^2 e^{-2\nu} \left(\mu'+\nu'\right) +  \frac{l^2}{r^2 } \left(\mu' + \frac{1}{r} \right) + \mu' \right] ,
\eea 
the specific energy and the specific angular momentum on a circular orbit can be written in terms of the radius due to the conditions
\bea
\frac{dr}{d\ta} = 0, \quad \frac{d^2r}{d\ta^2} = 0.
\label{cond1}
\eea
Specifically, we find 
\bea
\ep^2 =  \frac{\mu' + \nu' - \frac{1}{r}}{\nu' - \frac{1}{r}} e^{2\nu} , \quad l^2 = \frac{\mu' r^2}{\nu' - \frac{1}{r}} .
\label{coel}
\eea

In an EMRI system, denoting $m_2$ as the smaller object, the energy of the system can be approximated as $E = \ep m_2$. As the circular orbit adiabatically decays, the energy change rate is therefore
\bea
\frac{dE}{dt} = m_2 \frac{d\ep}{dr} \frac{dr}{dt} .
\label{decay2}
\eea
Connecting Eq.~\rf{decay1} and Eq.~\rf{decay2}, and using the relation between $\ep$ and $r$ in Eq.~\rf{coel}, we can express the changing rate of $r$ in terms of $r$, thus solving the relation between $r$ and $t$ while the circular orbit slowly decays. With the change of $r$ with respect to $t$, the phase on the circular orbit is then calculated by integrating the angular velocity $\omega = d\ph/dt$ over $t$, namely
\bea
\ph = \int_0^t \frac{d\ph}{d\ta} \frac{d\ta}{dt} dt = \int_{r_0}^r \frac{l}{\ep} \frac{e^{2\nu}}{r^2} \frac{dt}{dr} dr ,
\eea
where $r_0$ is initial radius of the circular orbit.

We consider a prototype EMRI system as in Ref.~\cite{Maselli:2021men}, which consists of a SMBH with mass $m_1=10^6 \, M_\odot$ and a smaller BH with mass $m_2=10 \, M_\odot$.
We set the initial separation between the two BHs $r_0$ to be $5 R^S_{m_1}$, i.e.\ $5$ Schwarzschild radii of the SMBH. 

\begin{figure}
 \includegraphics[width=\linewidth]{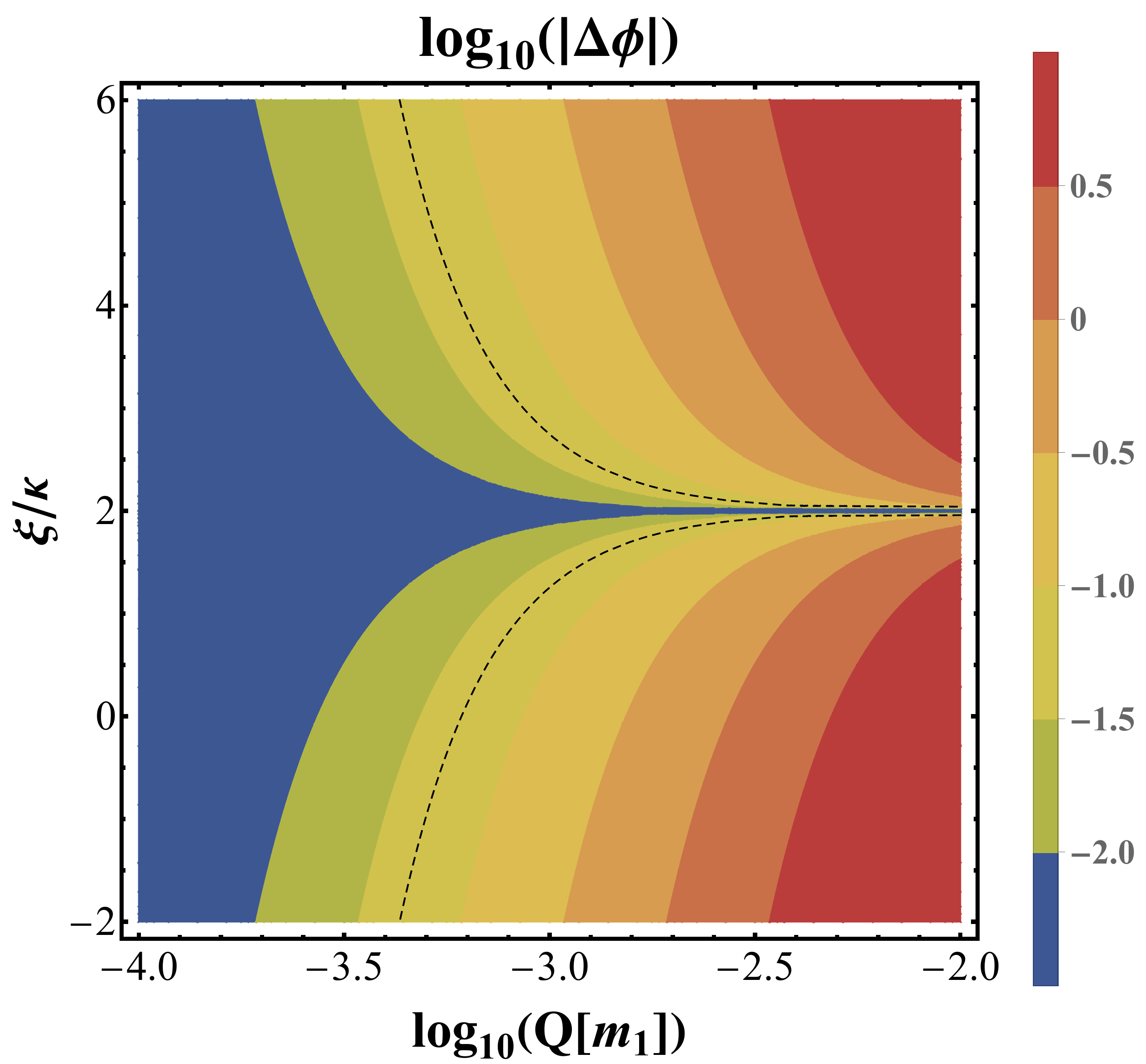}
 \caption{Logarithm of the accumulated orbital phase difference $|\Delta\phi|=|\phi_{\text{GR}}-\phi_{\text{Bum}}|$ between GR and bumblebee model for an EMRI after one-year evolution. The starting position for the smaller BH is $5R^S_{m_1}$. The black dashed lines denote $|\Delta\phi|=0.05$.  }
\label{deltaphi}
\end{figure}

For $\xi>2\kappa$, the orbital phase difference $\Delta\phi=\phi_{\text{GR}}-\phi_{\text{Bum}}$ is always negative when $Q\neq0$, which means the existence of the vector charge will accelerate the inspiral of the system. While for $\xi<2\kappa$, the inspiral will slow down when the SMBH carries a vector charge.
In Fig.~\ref{deltaphi}, we show the absolute value of the phase difference in logarithm after an assumed a one-year evolution.

Following Refs.~\cite{Lindblom:2008cm,Bonga:2019ycj}, we choose the GW phase difference to be $0.1$\,rad as the threshold for a dephasing observable by LISA for a system detected with signal-to-noise ratio (SNR) of $\sim30$.
As the frequency of GWs is twice the orbital frequency, this threshold corresponds to $|\Delta\phi|=0.05$ for the orbit, which is marked by the black dashed lines in Fig.~\ref{deltaphi}.
From this figure, we can estimate that we will be capable to probe the vector charge as small as $10^{-3}$ to $10^{-2}$ when $4\kappa \gtrsim |\xi-2\kappa|\gtrsim 0.2\kappa$.

\section{Waveform comparison}
\label{waveformcom}

With the orbital evolution of the EMRI system, we now can calculate the GW waveform up to quadrupole radiation using
\beq
\label{hijgr}
 h_{ij}=\frac{4G\eta M}{c^4 D_L}\left(v_iv_j-\frac{G m}{r}n_in_j\right),
\eeq
where $\bm{v}$ is the relative velocity and $\bm{n}$ is the direction of the separation vector of the two BHs.
To construct the two tensor polarizations, we adopt a new ``detector-adapted" coordinate system as in Ref.~\cite{Poisson:2014}, where the coordinate directions are given by 
\begin{align}
    \bm{e}_X &=(\cos\zeta, \, -\sin\zeta,\, 0), \nonumber \\
    \bm{e}_Y &=(\cos\iota\sin\zeta,\, \cos\iota\cos\zeta,\, -\sin\iota), \nonumber \\
    \bm{e}_Z &=(\sin\iota\sin\zeta,\, \sin\iota\cos\zeta,\, \cos\iota).
\end{align}
Here, $\iota$ is the inclination angle and $\zeta$ is the longitude of pericenter.
In this selected transverse basis, the transverse-traceless tensor polarizations are given by \cite{Poisson:2014}
\begin{align}
h_+=& -\frac{2\eta}{c^4 D_L} \frac{(GM)^2}{r}  (1+\cos^2 \iota) \cos(2\phi +2\zeta) ,
\nonumber \\
h_\times=& -\frac{4\eta}{c^4 D_L} \frac{(GM)^2}{r} \cos\iota \sin(2\phi +2\zeta) .
\end{align}


\begin{figure*}
 \includegraphics[width=\linewidth]{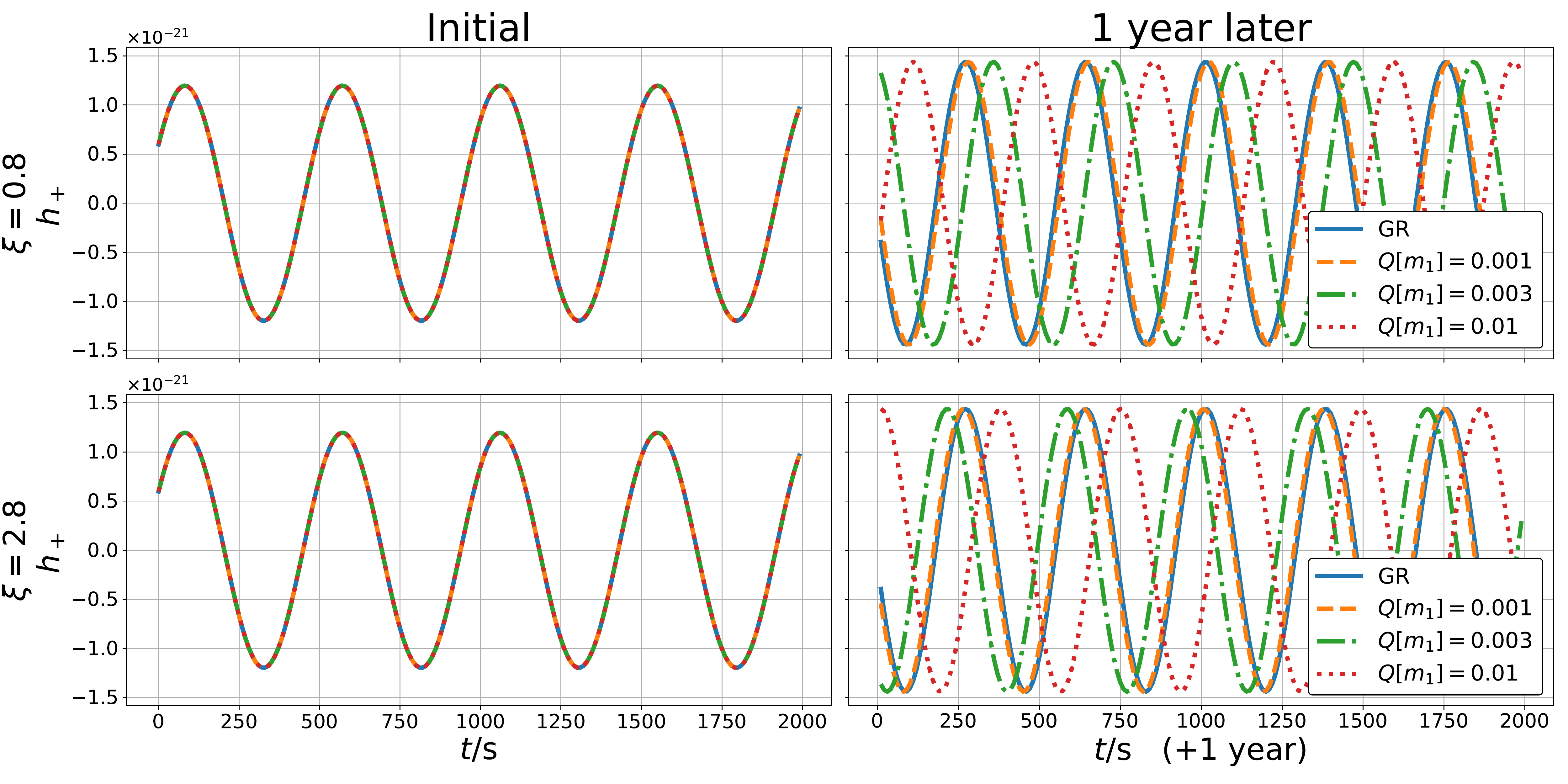}
 \caption{Waveform comparison between GR and bumblebee gravity with different vector charges. We show the case where $\xi=0.8$ and $\xi=2.8$ at the first and second row respectively. The waveform in GR is shown in solid line, while the waveforms in bumblebee theory with different vector charges are shown in dashed, dash-dotted and dotted lines respectively.}
\label{waveform}
\end{figure*}


In Fig.~\ref{waveform}, as an example we show the waveforms of the plus mode for GR and bumblebee gravity at the beginning of the evolution and after one year.
Here we adopt $\iota=\zeta=\pi/3$ and a luminosity distance $D_L=100\,$Mpc.
Initially, the smaller BHs start at the same position $10R^S_{m_1}$, and the waveforms overlap with each other in these two theories. 
While they become visually distinguishable after one-year accumulation of the orbital phase difference for $Q \gtrsim 0.003$ when $\xi=0.8\kappa$ or $2.8\kappa$.  

Instead of using low-frequency approximation, we consider the full frequency response of the detectors, where the antenna pattern function can be expressed as 
\begin{align}
F^A(t) = & \frac12 [\hat{u}_e^i \hat{u}_e^j T(f, \hat{\bm{u}}_e \cdot \hat{\bm{w}}_e) -\hat{v}_e^i \hat{v}_e^j T(f, \hat{\bm{v}}_e \cdot \hat{\bm{w}}_e) ]
\nonumber \\
& \times e^A_{ij}(\th_e,\phi_e,\psi_e) .
\end{align}
Here $A=+,\times$, $e_{ij}$ is the polarization tensor, $T$ is the transfer function \cite{Estabrook1975,Cornish:2001qi}. The vectors $\hat{\bm{u}}_e$, and $\hat{\bm{v}}_e$ are the unit vectors along the directions of the two arms of the detector, $\hat{\bm{w}}_e$ is the propagation direction of GWs. The angles $\theta_e$ and $\phi_e$ denote the sky location of the source, and $\psi_e$ is the polarization angle. 
They are all measured in the ecliptic coordinate system and the details can be found in Refs.~\cite{Liang:2019pry, Liu:2020nwz}.
The signal $s(t)$ recorded in the detector can be expressed as 
\beq
\label{tdwaveform}
s(t) = F^+(t) h_+(t) +F^\times(t) h_\times(t) .
\eeq
To include the Doppler modulation effects by the motion of LISA, we further modify the phase of the waveform as $\Phi(t)\to \Phi(t)+\Phi'(t) R_{\text{AU}} \sin \theta_e \cos(2\pi t/T-\phi_e)$ \cite{Maselli:2021men}. Here, $R_{\text{AU}}$ is the astronomical unit and $T=1$ year.
The noise-weighted inner product is defined as
\beq
\langle s_1|s_2 \rangle=2\int_{f_{\rm min}}^{f_{\rm max}}\frac{\tilde{s}_1(f)\tilde{s}_2^*(f)+\tilde{s}_1^*(f)\tilde{s}_2(f)}{S_{n}(f)}df,
\eeq
where $\tilde s(f)$ is the Fourier transformation of $s(t)$, and $S_n(f)$ is the power spectral density of LISA \cite{Robson:2018ifk}.
The SNR $\rho$ of signal $s$ is just simply $\rho=\sqrt{\langle s|s \rangle}$.
To quantitatively measure how much two signals differ from each other, we calculate the faithfulness $\mathcal{F}$, which is defined as
\beq
\label{eqfaith}
\mathcal{F}[s_1,s_2]=\max_{\{t_c,\phi_c\}}\frac{\langle s_1\vert
	s_2\rangle}{\sqrt{\langle s_1\vert s_1\rangle\langle s_2\vert s_2\rangle}}\ ,
\eeq
where $\{t_c,\phi_c\}$ are time and phase offsets \cite{Lindblom:2008cm}.
For a model with a parameter dimension $d$, if the faithfulness $\mathcal{F}$ is smaller than $\sim 1-d/(2\rho^2)$, it means that the two waveforms are significantly different and are distinguishable for the detectors \cite{Flanagan:1997kp,Lindblom:2008cm}.
In our model, the time-domain waveform Eq.~(\ref{tdwaveform}) is determined by ten parameters $\vec{\theta}= \{M,\eta,D_L,\theta_e,\phi_e,\iota,\zeta,r_0,\Phi_0,Q\}$ for the bumblebee model and nine for GR.
Following Ref.~\cite{Maselli:2021men}, we consider the last one-year evolution of the EMRI system before plunge and rescale the luminosity distance to make $\rho$ to be 30. Therefore, the threshold faithfulness in our study is $\mathcal{F}_{\text{th}}=0.989$.

\begin{figure}
 \includegraphics[width=\linewidth]{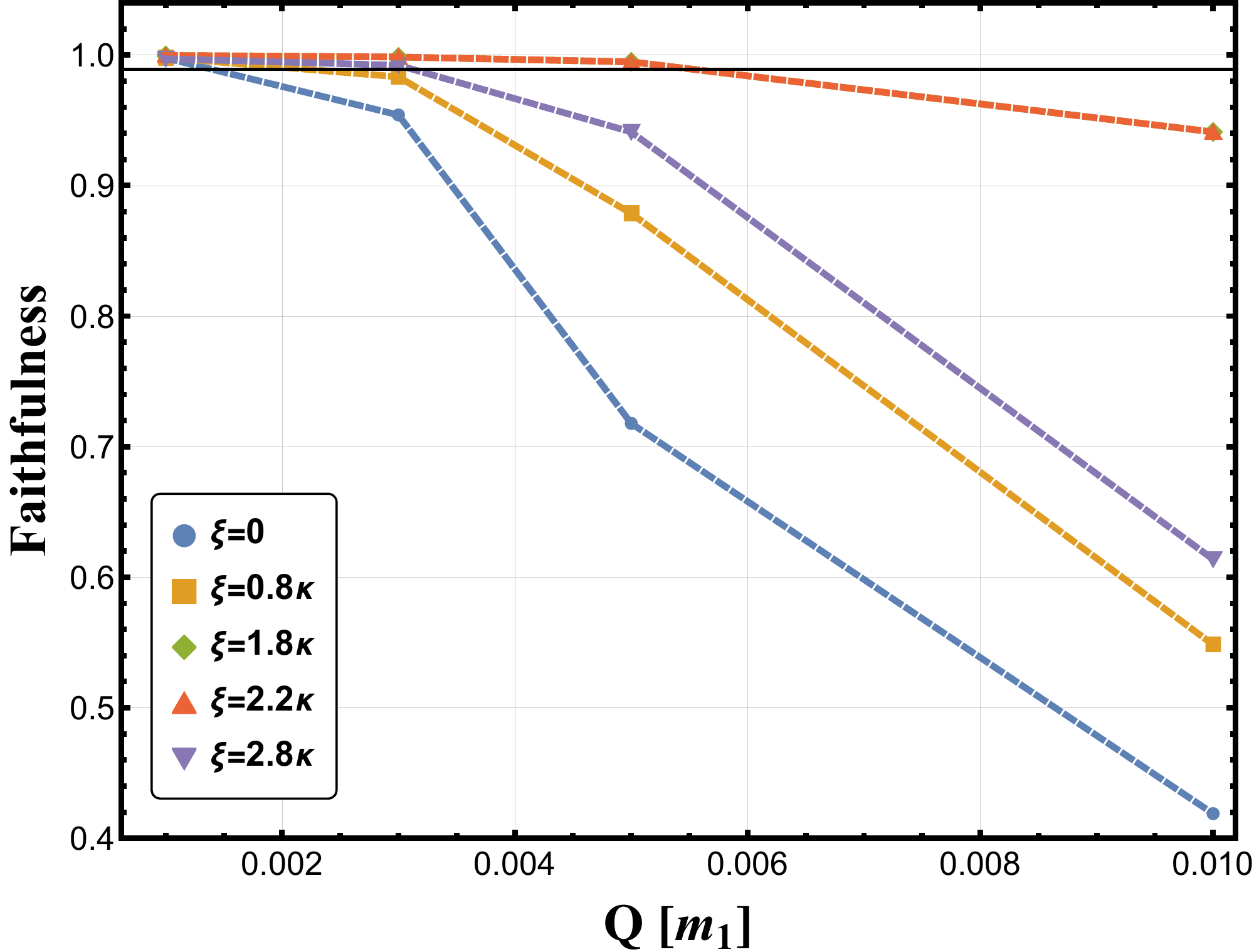}
 \caption{Faithfulness between the waveforms computed in GR and the bumblebee model. The threshold $\mathcal{F}_{\textbf{th}}=0.989$ is denoted by the thin black line. }
\label{faithfulness}
\end{figure}

The faithfulness between waveforms in GR and bumblebee model is shown in Fig.~\ref{faithfulness}.
On one hand, with the increase of the vector charge, the waveform in the bumblebee model deviates more from that in GR, thus the faithfulness decreases as expected. 
On the other hand, when $|\xi-2\kappa|$ becomes larger, we also have smaller faithfulness for waveforms in GR and the bumblebee model.
For $Q=0.001$, we have $\mathcal{F}>0.997$ for all the five cases we consider.
For $Q=0.003$, we have $\mathcal{F}=0.954$ when $\xi=0$ and $\mathcal{F}=0.983$ when $\xi=0.8$.
For $Q=0.005$, we have $\mathcal{F}=0.941$ when $\xi=2.8$. 
For $Q=0.01$, we have $\mathcal{F}=0.941$ when $\xi=1.8$ and $\xi=2.2$.
We conclude that with a one-year observation of the EMRI system, we can probe the vector charge as small as $Q\sim \mathcal{O}(10^{-3})-\mathcal{O}(10^{-2})$ when $|\xi-2\kappa|\gsim 0.2\kappa$.

On one hand, if the source is closer to us or we have a detector network, we will have a  higher SNR. Then we have higher threshold for the faithfulness and we can distinguish the waveforms in GR and the bumblebee model with a smaller vector charge. 
On the other hand, with a longer observation time which is like the case in reality, the accumulated phase difference between two theories becomes larger, and it also helps us probe a smaller vector charge.

\section{Discussion}
\label{discussion}

In the bumblebee model, the BH solution reduces to Schwarzschild BH when the coupling constant $\xi=2\kappa$. When $\xi\neq 2\kappa$, the deviation from Schwarzschild metric becomes prominent when the BH carries a large vector charge (see Fig.~\ref{rh}).
Such a vector charge is induced by the bumblebee field, and is independent of the spin and mass of the BH. Thus, it can be considered as a hairy BH solution in this theory.

In our previous study \cite{Xu:2022frb}, we used the EHT observations to constrain the vector charge and we found the constraints quite weak. 
As was showed in Fig.~10 in Ref.~\cite{Xu:2022frb}, current constraint on the vector charge carried by the Sgr~A* BH is $Q\lesssim \mathcal{O}(1)$ for $\xi=3\kappa$, $\kappa$ or $-\kappa$.
In this paper, we use an EMRI system with the inspiraling stellar-mass BH as a probe to see how well we can detect the deviation from the Schwarzschild metric, namely, that how well we can distinguish the Schwarzschild BH from the hairy BH.

We first calculated the orbital evolution of the EMRIs both in GR and in the bumblebee model, and estimated the orbital phase difference with various $\xi$ and $Q$, as shown in Fig.~\ref{deltaphi}.
Due to the accumulated orbital phase difference, the difference between the waveforms in these two theories also become more and more significant as the orbit evolves. In addition to visualizing the difference of waveforms in Fig.~\ref{waveform}, we also calculated the faithfulness between the waveforms to quantify the difference and its detectability.  

Combining the information from the orbital phase difference and faithsfulness between the waveforms, we find that with one-year observation of an EMRI, we can probe the vector charge as small as $Q\sim \mathcal{O}(10^{-3})$ , which is about $3$ orders of magnitude smaller than that with the EHT. 
It is not surprising, since when considering EMRIs, we have made use of a larger vicinity of the strong-field region of the SMBH, and the long time inspiral in this region ensures the phase difference accumulation to be significant. 
Since the GW observations are very sensitive to the phase, we anticipate that EMRIs can help us probe a much smaller vector charge.
In Refs.~\cite{Burton:2020wnj,Zi:2022hcc}, the threshold value of the electric charge of a SMBH for distinguishing it from a neutral SMBH is $\sim 10^{-3}$.
Considering that the bumblebee BH with vector charge is an extension of BH to the electric charge, our constraint is within expectation.

The limitation of the waveform model built here is the omission of the scalar and vector GW modes as well as tensor multipoles higher than the quadrupole. 
They all contribute higher-order corrections to the energy loss of the system.
To obtain the accurate waveforms of EMRIs, it is general to implement the Teukolsky formalism \cite{Teukolsky:1973ha} to calculate the loss rate of energy and angular momentum \cite{Cutler:1994pb}.
But it is theoretically complicated and computationally expensive to apply the Teukolsky formalism in the bumblebee gravity model, the reasons are listed as follows:
(i) As is shown in Ref.~\cite{Burton:2020wnj}, the master equations for the odd- and even-parity perturbations are complex in the Einstein-Maxwell theory, thus it is not surprising that they will be even more sophisticated when we consider the nonminimal coupling between the vector field and the tensor field.  We leave it for a future study;
(ii) Teukolsky-based waveforms themselves are computationally expensive to generate since they require the numerical integration of the Teukolsky equation and summation over a large number of multipole modes;
(iii) Currently we do not have analytic formulas for the general BH solution in the bumblebee model, instead, the solution is only available numerically. We need high requirements for the accuracy of the computation to avoid accumulated errors, which makes the computation even more costly.
Thus, we only take a very first step here to give an estimation of the detectability of the vector hair, and there is still a long way to go to provide accurate waveform templates for EMRIs in the bumblebee gravity.
When considering the extra radiation channel, it will accelerate the inspiral of the binary system. As we analyzed in  Sec.~\ref{orbit}, due to the modification on the metric, the inspiral is slower in the bumblebee model than that in GR when $\xi<2\kappa$ while it is faster when $\xi>2\kappa$. From this perspective, the constraint on the vector charge obtained in this paper is optimistic when $\xi<2\kappa$ and conservative when $\xi>2\kappa$.

Another direction to improve the usability of the waveform model for the unprecedentedly accurate data to be obtained by future space-based GW detectors is adding in corrections due to the rotation of the SMBH.
While the Kerr-like BH solution in the bumblebee model found in Ref.~\cite{Ding:2019mal} was pointed out to be incorrect \cite{Maluf:2022knd}, the slow-rotation approximation of the rotating BH in the bumblebee model was obtained in Ref.~\cite{Liu:2022dcn}. Extending our analysis to EMRI with a rotating SMBH in the bumblebee model is future work worth a detailed investigation.

\acknowledgments 
This work was supported by the National Natural Science Foundation of China (No. 12147120, No. 11975027, No. 11991053, No. 12247128), the China Postdoctoral Science Foundation (No. 2021TQ0018),
the National SKA Program of China (No. 2020SKA0120300), 
the Max Planck Partner Group Program funded by the Max Planck Society, 
and the High-Performance Computing Platform of Peking University.
R. X. is supported by the Boya postdoctoral fellowship at Peking University.


%

\end{document}